\documentclass[twocolumn,amsmath,amssymb,letterpaper,showpacs,prl]{revtex4}
\usepackage[tight]{subfigure}
\usepackage[dvips]{graphicx}
\usepackage{bm}
\usepackage{dcolumn}

\begin{document}


\title{Onset of Irreversibility in Cyclic Shear of Granular Packings}
\author{Steven Slotterback$^1$}
\author{Mitch Mailman$^1$}
\author{Krisztian Ronaszegi$^1$}
\author{Martin van Hecke$^2$}
\author{Michelle Girvan$^1$}
\author{Wolfgang Losert$^1$}
   \email{wlosert@umd.edu}
\affiliation{$^1$Department of Physics, and IREAP, University of
Maryland, College Park, Maryland, 20742}
\affiliation{$^2$Kamerlingh Onnes Laboratory, Universiteit Leiden,
PObox 9504, 2300RA Leiden, the Netherlands}
\date{\today}


\begin{abstract}
We investigate the onset of irreversibility in a dense granular medium subjected to cyclic shear in a split-bottom geometry. To probe the micro and mesoscale, we image bead trajectories in 3D throughout a series of shear strain oscillations. Though beads lose and regain contact with neighbors during a cycle, the global topology of the contact network exhibits reversible properties for small oscillation amplitudes. With increasing reversal amplitude a transition to an irreversible diffusive regime occurs.
\end{abstract}

\maketitle

The transition from reversible to irreversible dynamics in systems of many interacting particles is of fundamental importance to many-body physics. Granular materials under repeated shear deformations constitute excellent experimental systems with which to study such reversibility for out-of-equilibrium systems.  Since grains are macroscopic in size, thermal fluctuations are negligible \cite{DuranBook}, with particle trajectories determined by contact forces between neighboring grains. In ground breaking work on low concentration granular suspensions, Cort\'e and coworkers observed that particle collisions tend to disrupt the inherent reversibility of low Reynolds number fluid flow, and that reversibility is lost at strains and densities that allow a particle to collide with two or more neighbors \cite{2008NatPh...4..420C}. As a consequence, the onset of irreversibility decreases approximately as the square of particle density \cite{2005Natur.438..997P}.

For dense particle systems close to jamming, the threshold for exact spatial reversibility would thus become extremely small.  However, dense configurations of particles have additional constraints on their positions: the neighborhood of nearby grains in the local environment of the particle confine its position, providing a new reference for the motion of the particle.  This perspective opens up the possibility of an alternative form of reversibility with respect to the local neighborhood of particles.  Under shear strain, neighbors are lost, and correlations are important: rearrangements in dense configurations are highly collective and are known to exhibit a range of dynamical features, from local reversibility as found in foams \cite{PhysRevE.77.041505} to larger scale dynamical heterogeneities as found in 2D bidisperse cylinders \cite{PhysRevLett.95.265701,PhysRevLett.94.015701,PhysRevLett.102.088001}.  

In this article, we describe experiments that explore strains for which these rearrangements first emerge at scales larger than a grain diameter.  Since we probe extremely slow strain rates, the granular temperature is virtually zero and rearrangements are due to plastic deformations as opposed to random motion.  We characterize the local environment around particles, and the plastic deformations, through the analysis of the network of nearby particles.  In our experiments, the full three dimensional motion of beads within a split-bottom shear cell undergo cyclic shear. Dense granular systems under cyclic shear in two and three dimensions have been studied previously \cite{PhysRevE.81.060301,zhang:553,PhysRevLett.95.265701,losert04,PhysRevE.71.031301}, but the trajectories of all beads in the bulk have not been probed, and collective bead dynamics and reversibility are not well understood. 

We find that Mean Square Displacements (MSD) of beads indicate a crossover between two different behaviors and that a characteristic strain amplitude is associated with this crossover. In order to relate this change in behavior to changes in the neighbor network, we analyze neighbor changing events through the use of the Broken Link Network \cite{PhysRevE.83.061303}, which assigns a link to pairs of particles that then later move apart, under shear strain.  For a strain  amplitudes below the aforementioned characteristic amplitude, the largest interconnected component of the BL network nearly disappears after shear reversal, while above the characteristic amplitude a giant component appears to emerge and survive the shear reversal.   We posit a novel type of reversibility/irreversibility transition under cyclic shear strain: one that distinguishes between a regime where rearrangements are essentially local and random, never leading to a qualitative change in the topology of the BL network, and a regime where the topology is modified significantly by the formation of a giant component.  We refer to this phenomenon as 'topological irreversibility,' by virtue of the fact that the topology of the BL network is being irreversibly altered by the formation of a giant component.

\begin{figure}[t,h]
\centering
\DeclareGraphicsRule{*}{eps}{*}{}
\includegraphics*[width=0.48\textwidth]{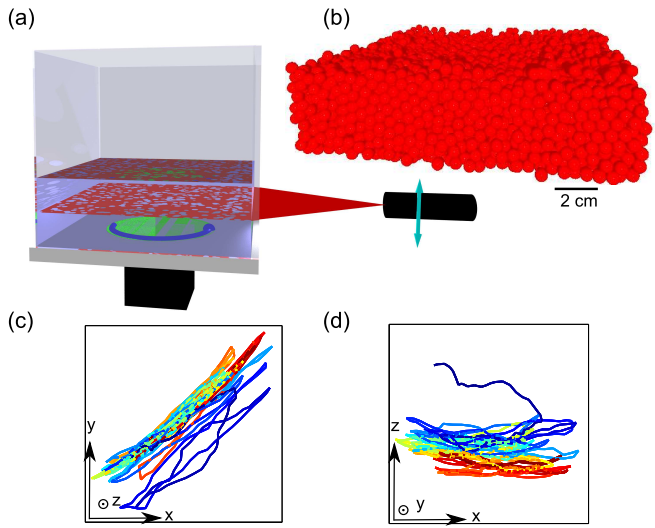}
\caption{(a) Split-bottom shear cell with acrylic beads in index matched, dyed fluid.  A laser sheet illuminates cross-sections for imaging. (b) As shown in this reconstruction, the position of all beads is found by scanning the laser sheet sheet across the whole container and automatic particle recognition. (c)-(d) A typical trajectory of a bead under oscillatory shear strain with a reversal amplitude of $40^{\circ}$ from (c) above and (d) the side (cube sides are 2 bead diameters, color changes from blue to red color with increasing time). } 
\label{fig:setup}
\end{figure}

\paragraph{Setup ---}  To study motion of all beads, we built a modified version of the split-bottom geometry \cite{PhysRevLett.92.094301,fenistein03} that allows for 3D imaging. In this geometry, shear is applied from below by rotating a disk mounted flush with the bottom. A shear zone is then emanating from the edge of a rotating disk --- much of the shearzone is far from boundaries, the shear zone is wide and smooth, and nothing obstructs its view, making it ideally suited for 3D observations of grain flows with smooth gradients.

We use a square box with a 15 cm by 15 cm bottom and a 9 cm diameter disk at the bottom.  In order to provide traction for the beads above, we attach a sparse coating of 3 mm beads to the bottom of the container and disk surface. We fill the container with 3/16'' diameter acrylic beads. 
We immerse the beads in an index matched fluid (Triton X-100).  A fluorescent dye (Nile Blue 690 Perchlorate) is added to the fluid, along with a small amount of HCl (1.0 mL per 1.0 L of fluid) to stabilize the mixture.

We use the 3D imaging technique described in~\cite{PhysRevLett.101.258001} and illustrated in Fig.~\ref{fig:setup}, but with an improved bead finding algorithm.  We use a 3D version of the convolution described in~\cite{PhysRevE.70.031303} to find the position of each bead to better than half a pixel, or 4\% of a bead diameter. We then track the motion of individual beads with a tracking algorithm developed by Grier et al.~\cite{crocker96} to find trajectories such as the one in Fig. \ref{fig:setup}c-d. With the full trajectories for about 98\% of beads in our system outside the bottom layer, we can now analyze how beads move with respect to their neighbors. 

We choose to fill the container to a filling height $H$ equal to the radius of the shearing disk $R_s$.  This regime, which has only been approached experimentally in a limited number of studies \cite{PhysRevLett.96.038001,PhysRevE.83.061303,PhysRevLett.96.118001}, is between two predicted regimes wherein the shear bands in a constant strain angular velocity profile transition from annular regions that widen above the disk edge ($H/R_s < 0.7$) to a dome-shaped region above the disk ($H/R_s >> 1$) \cite{PhysRevLett.92.214301}.  We choose this intermediate regime in order to maximize the fraction of grains undergoing differential flow.

The angular velocity flow field is quite heterogeneous in the case of constant shear.  The differential flow, as shown in \cite{PhysRevE.83.061303} is greatest in a region above the shearing disk.  We choose to limit our analysis to this region of maximum relative strain.  We find this region by calculating the mean angular velocity in annular regions about the axis of disk rotation.  We disregard axial and radial velocities, since the bulk remains stationary on average.  We use these values to find the approximate differential strain in the axial and radial directions.  The regions with relative strains in the lower 60th percentile are excluded.

Imaging cannot distinguish whether two hard spheres are in contact or just close together; instead we choose a maximum separation criterion to identify the nearest neighbors of each bead based on a preference for tangential motion of neighboring beads~\cite{PhysRevE.83.061303,2009AIPC.1145..489S}.  Based on this criterion, we consider beads whose centers are within $\approx 1.08$ bead diameters of one another to be nearest neighbors, since beads that are separated by greater distances show no such preference for tangential relative motion.  Thus, the beads have an effective radius $R = 0.54$ bead diameters.

{\em Protocol ---} To create reproducible initial conditions representative of steady shear, we rotate the disk two full
revolutions at 1 mrad/sec.  At this rate of strain, the granular flow in our fluid immersed system is independent of strain rate,
with the same velocity profile as a dry granular flow~\cite{PhysRevE.82.060301}. We then reverse the direction of
rotation of the disk and move the disk for a strain of $\theta_r$ ($=2^{\circ},4^{\circ},10^{\circ},20^{\circ},40^{\circ}$).  We
reverse the direction again and rotate the disk by $-\theta_r$ to its original position, thus straining and reversing the system in a cyclic fashion.  We perform 20 successive cycles in each experiment, taking a 3D image of the system after every
$2^{\circ}$ of strain.

\begin{figure}[t]
\centering
\includegraphics[width=0.52\textwidth]{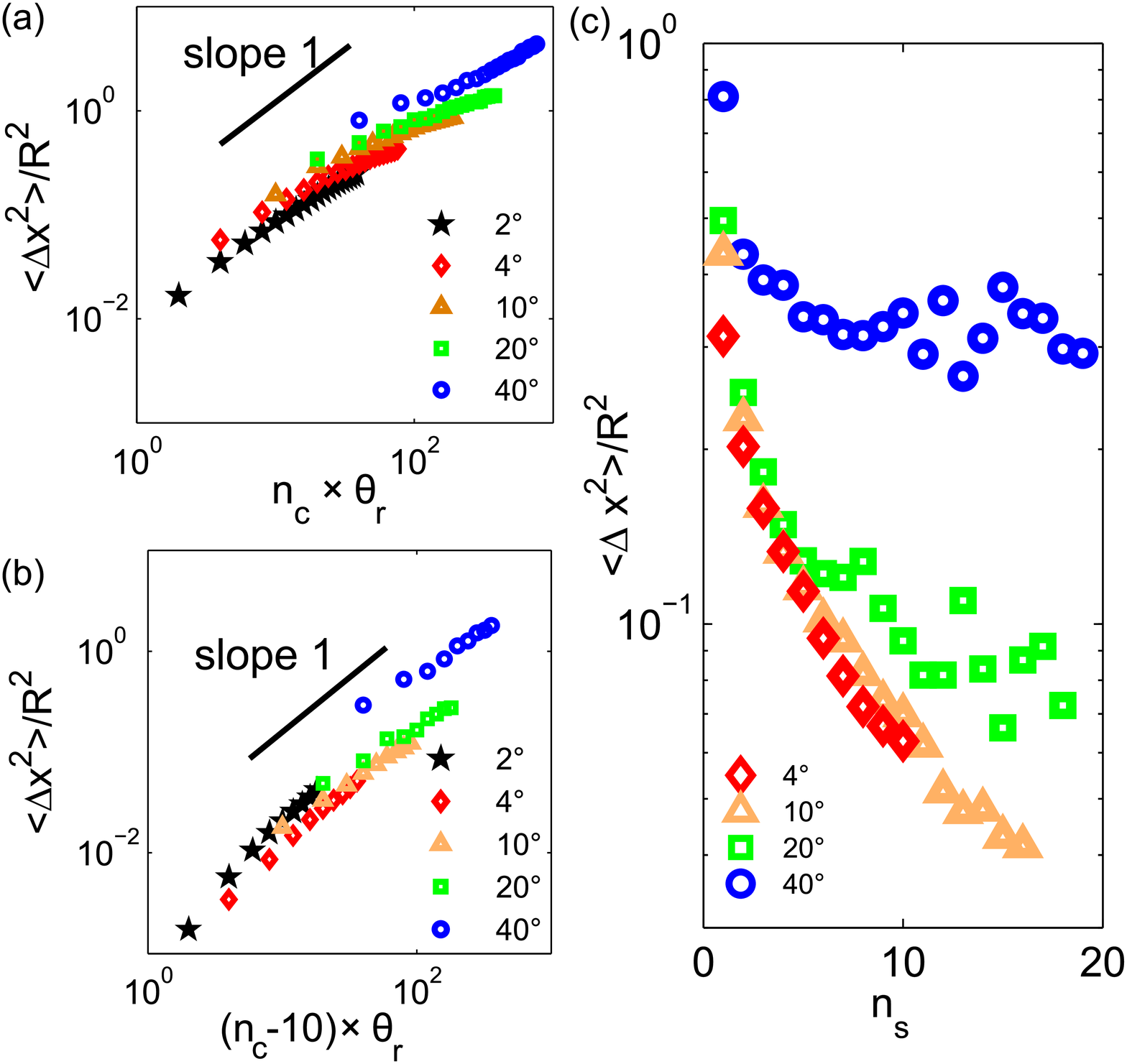}
\caption{(a)-(b) Mean Square Displacements (MSDs) at the end of
each cycle as a function of cycle number for different cycle
amplitudes, scaled by the effective bead radius $R$: (a) Reference frame at the start of the first cycle,
(b) The start of the 10th cycle is chosen as a reference frame.
(c) MSD after $80^{\circ}$ of absolute strain vs. reference cycle
number. } \label{fig:msdsall}
\end{figure}

\paragraph{Mean Square Displacements ---}
The first measure of reversibility we employ is to test how far a bead has moved after a full cycle. The mean square displacements
(MSD) of the beads in the shear zone at the end of each of the 20 cycles are shown as function of cycle number, $n_c$, in
Fig.~\ref{fig:msdsall}a. For all but the largest reversal amplitude $\theta_r$, the slope of the MSD decreases with increasing cycle number to slopes below 1, signaling sub-diffusive behavior.

To explore whether the system becomes more reversible after several shear cycles, we choose the beginning frame of a later starting cycle $n_s$ as a reference frame. Fig.~\ref{fig:msdsall}b illustrates that for $n_s=10$, the MSD is up to an order of magnitude smaller, suggesting that bead positions in this cyclic shear flow become more reversible after several cycles. Strikingly, the MSD for the largest reversal amplitude ($\theta_r=40^{\circ}$) is considerably larger in magnitude.

To compare the rearrangements after one cycle of amplitude $\theta_r$ with two cycles of amplitude $\theta_r/2$ etc, we
calculate the MSD after a fixed absolute strain $\Delta \theta$ (relative to a chosen reference frame) which we fix here at $80^{\circ}$. Fig.~\ref{fig:msdsall}c shows the MSD at this fixed total strain as function of the start cycle $n_s$. We see that for $\theta_r = 40^{\circ}$, the magnitude of the MSD does not depend strongly on the choice of reference cycle after a short transient of about 3 cycles, while the MSD for smaller $\theta_r$ continue a downward trend for all cycles investigated.

In terms of spatial reversibility where all beads return to their original coordinates, all three probes of the MSD strongly suggest a qualitative change in behavior between $\theta_r=20^{\circ}$ and $\theta_r=40^{\circ}$: for large amplitude, the motion is irreversible, and reaches a steady, diffusive state, while for small amplitudes, the cyclic shear strain process becomes increasingly reversible (i.e. decreasing MSD magnitudes) and the system is aging.

\begin{figure}[t]
\centering
\includegraphics[width=0.48\textwidth]{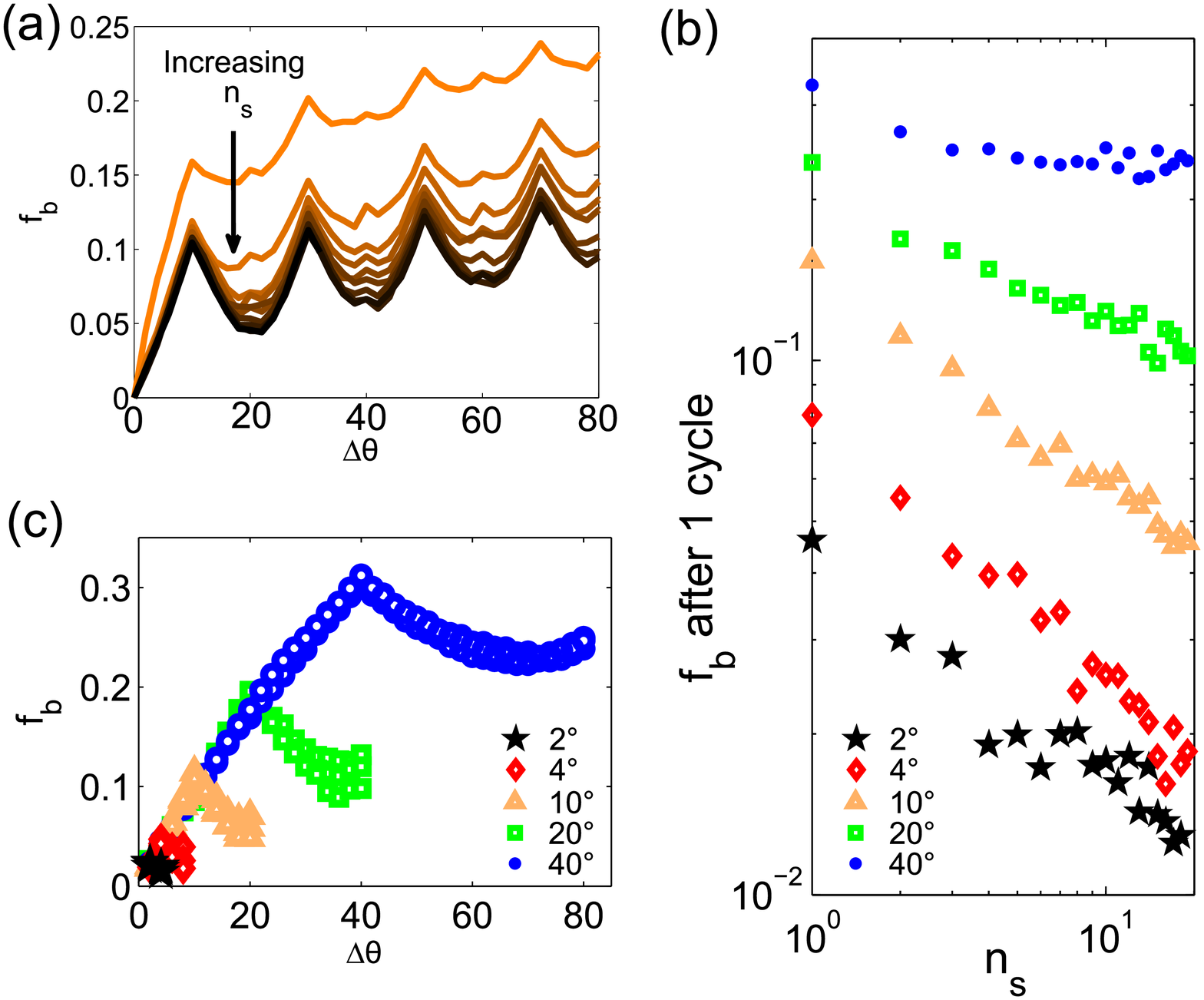}
\caption{(a) Fraction of links broken, $f_b$ for shear reversal vs.
$2^{\circ}$ strain step for the case of $\theta_r = 10^{\circ}$. (b)
$f_b$ after one cycle as a function of the starting cycle $n_s$, for all $\theta_r$. (c) $f_b$ as a function of $\Delta\theta$ for representative curves at $n_s$=5, 10 and 15.}
\label{fig:PersistenceTrends}
\end{figure}

{\em Broken Links ---} As a second measure of reversibility we probe the network of nearest neighbors. We first identify all
neighbors in a reference frame at the start of a cycle.  Then we calculate, as function of frame number, the fraction of links broken $f_b$, defined as the fraction of neighbor pairs from the reference frame that are no longer within the cutoff distance used to define neighbors. The approach of~\cite{PhysRevE.83.061303} has been modified here to include the possibility of ``healing.''  A link is healed when a neighbor pair from the reference frame that has been broken later reforms.  A healed link does not appear in the broken link network.  Note that other pairs of beads may become neighbors in later frames but this is not considered in the broken link analysis.

In Fig.~\ref{fig:PersistenceTrends}a we show the evolution of $f_b$ as a function of $\Delta \theta$, for $\theta_r = 10^{\circ}$
--- other $\theta_r$ show comparable results.  The top curve shows $f_b$ with $n_s$=1. We observe a strong decrease in $f_b$ when
the strain direction is reversed (at $10^{\circ}$) --- broken links reform after reversal. Additionally, fewer additional links are broken in each subsequent cycle --- the motion becomes less irreversible for later times. The ten lower curves in
Fig.~\ref{fig:PersistenceTrends}a show $f_b$ when the start of a later cycle is chosen as a reference frame, illustrating that the
recovery of neighbors after a full shear strain cycle (at arrow) improves with increasing cycle number.

In Fig.~\ref{fig:PersistenceTrends}b we show $f_b$ after one cycle as a function of the starting cycle $n_s$, which provides a local measure of the reversibility of the configuration after one cycle.  Clearly, the fraction of links broken decreases with increasing $n_s$ (indicating an increase in reversibility) for all but the largest reversal amplitude, which appears to level off.  As a simple local metric, $f_b$ captures what fraction of the neighbor network is evolving with strain, a topological analog to the ``particle activity" of \cite{2008NatPh...4..420C}.  Indeed, the apparent power dependence of the decay on $n_s$ in the reversible regime, and a leveling off for the highest reversal amplitude, is consistent with the observation of a reversible-irreversible transition observed in  \cite{2008NatPh...4..420C}.  However, as illustrated in Fig.~\ref{fig:PersistenceTrends}c, $f_b$ cannot identify this local reversibility in the neighbor network over the period of one cycle  - $f_b$ after one cycle increases approximately linearly with cycle amplitude without a clear change in the form of the dependence on $\Delta\theta$.

\begin{figure}[t]
\centering
\subfigure[]{\includegraphics[width=0.42\textwidth]{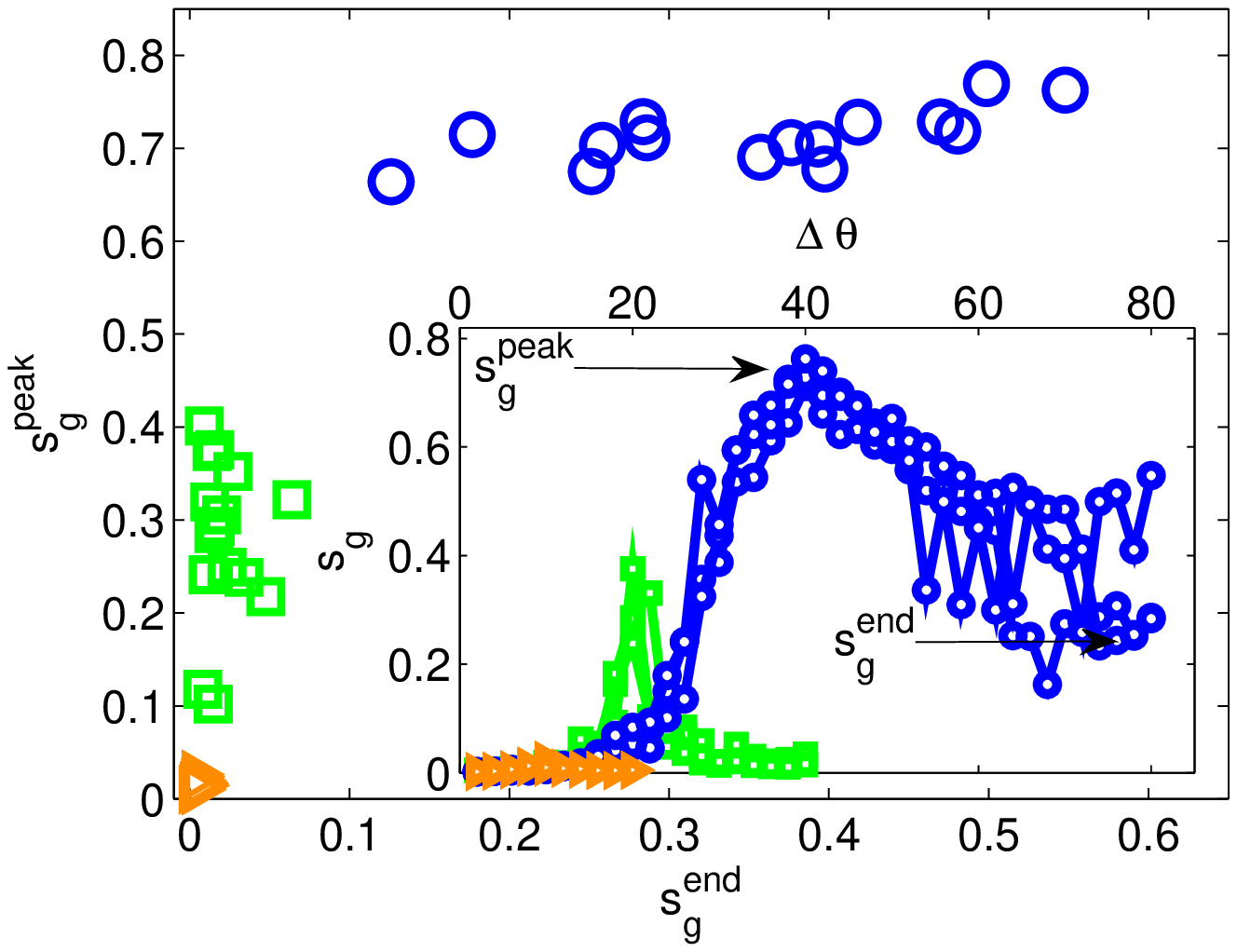}}
\subfigure[]{\includegraphics[width=0.42\textwidth]{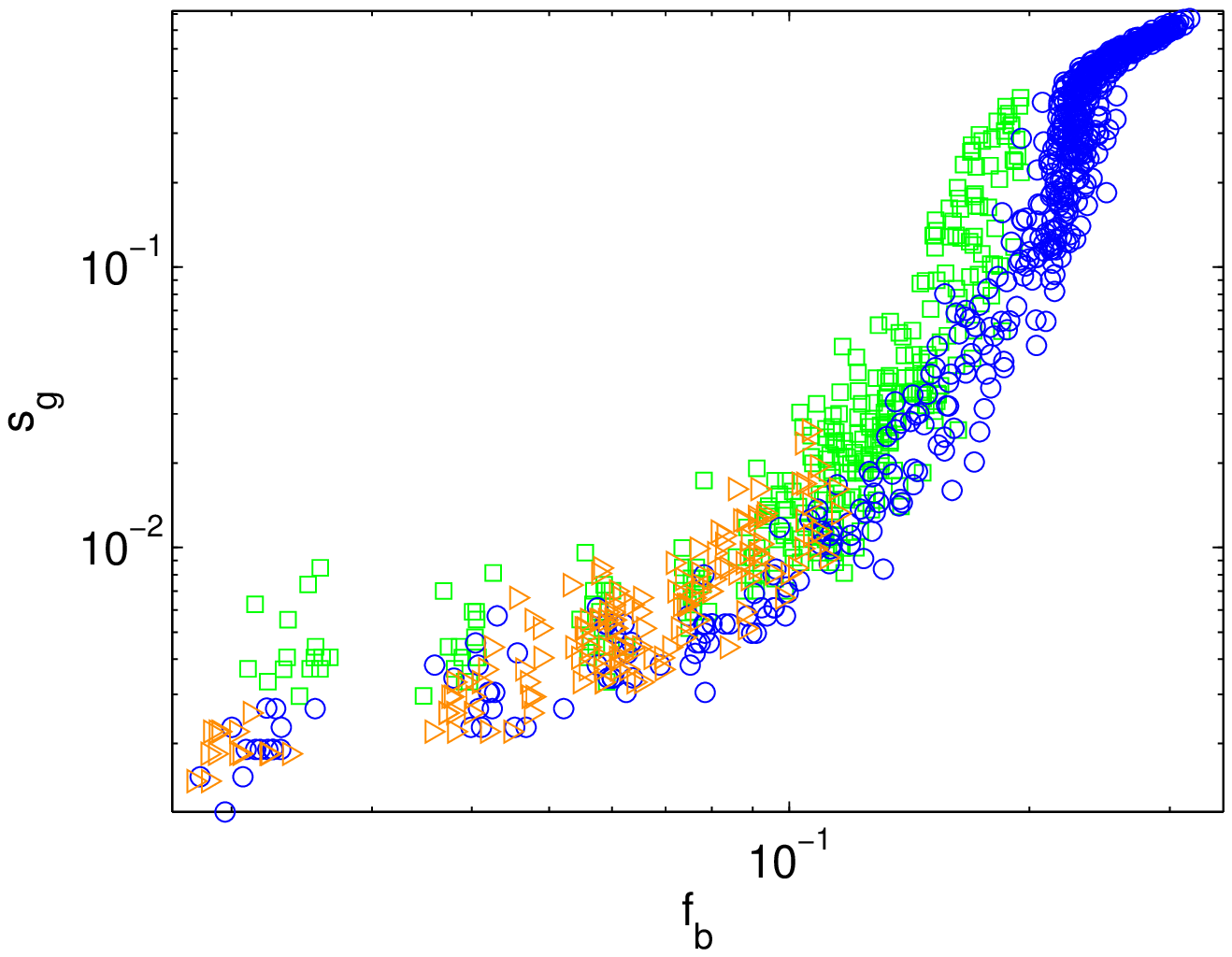}}
\caption{ (a) The peak largest component size versus the final largest component size, for each cycle. (a, Inset) $s_g$ as a function of $\Delta \theta$ over one cycle for $n_s=$5, 10, and 15. Peak and end values for $s_g$ are pointed to for the $40^{\circ}$ case. (b) $s_g$ as a
function of the fraction of links broken $f_b$ for each reference
network used, for $\theta_r$ of $10^{\circ}$(orange triangles),
$20^{\circ}$(green squares) and $40^{\circ}$(blue circles).}
\label{fig:giantcomponent}
\end{figure}

Our characterization of the irreversibility revealed a striking difference between the growth of mean square displacements and fraction of links broken with driving amplitude $\theta_r$. The MSD barely grows with $\theta_r$ for small driving amplitudes (see in particular Fig.~\ref{fig:msdsall}c for small start cycles, where all data for $\theta_r$ from $4^{\circ}$ to $20^{\circ}$ strongly overlap), but then exhibits a jump for $\theta_r =40^{\circ}$. In contrast, the fraction of links broken $f_b$ grows approximately linearly with $\theta_r$ (see Fig.~\ref{fig:PersistenceTrends}c) and so over a single strain cycle $f_b$ fails to distinguish $\theta_r=40$ from the smaller reversal amplitudes. 

We believe that this difference occurs along with the percolation of the network of links broken. In this picture, a small fraction of links broken leads to small clusters of failures that upon reversal are mostly healed.  Consequently, after a cycle the beads return to their starting position, leading to a small
MSD. 
However, once the fraction of links broken is sufficiently large, a dominant failing cluster forms.  Upon reversal, this dominant failing cluster is not fully healed, and after one cycle, beads end up with different neighbors and at different locations, leading to a large MSD.

To probe this picture, we investigate how links that break are connected with each other by observing an aspect of the topology of the broken links network. Under steady shear, it is known that the size of the largest component, i.e, the largest cluster of beads exhibiting inter-connected broken links, shows a transition beyond a characteristic strain (see~\cite{PhysRevE.83.061303}). Here we adapt this measure to oscillatory shear strain, and follow $s_g$, which is the fraction of beads in the largest component, as function of absolute strain $\Delta \theta$, with reference frames chosen at the beginning of each cycle.

Fig.~\ref{fig:giantcomponent} provides qualitative support for our picture.  In the inset of Fig.~\ref{fig:giantcomponent}a we show that for $\theta_r \leq 20^{\circ}$, the size of the largest component $s_g$ initially grows with $\Delta \theta$ but then shrinks back to almost zero when the motion is reversed (links are reformed after reversal), while for $\theta_r = 40^{\circ}$, $s_g$ remains substantial after completing a full cycle --- hence links remain broken. To stress this qualitative difference, we collect the data for the last 15 cycles, characterize the functional form of $s_g(\Delta \theta)$ by its peak size at the reversal, $s_g^{peak}$, and the largest component size at the end of the cycle, $s_g^{end}$, and make a scatter plot of $s_g^{peak}$ versus $s_g^{end}$ (Fig.~\ref{fig:giantcomponent}a). This reveals a clear separation of the $10^{\circ}$ and $20^{\circ}$ data which form peak component sizes that can be almost fully reduced to zero, from the largest amplitude $40^{\circ}$ where broken links interconnect a significant part of the system at the end of a cycle.

In Fig.~\ref{fig:giantcomponent}b we show scatter plots of $s_g$ versus $f_b$ for three cycle amplitudes, which show that for each
cycle amplitude, the data collapses onto a master curve quantitatively similar to that of \cite{PhysRevE.83.061303}.  For the largest driving amplitude, a regime emerges where $s_g$ has a dependence on $f_b$ that is consistent with a power law --- a collective, percolation effect.  

\paragraph{Discussion ---}
We have studied a granular system under shear reversal at microscopic and mesoscopic scales. Such dense systems are never found to be strictly reversible, but we have shown that irreversibility comes in two distinct flavors: spatial and topological. For small driving amplitudes, neighbor links are broken but most reform, there are no large clusters, and the mean square displacements are small, and importantly, the cycle-to-cycle mean square displacements decrease with cycle
number: the system is aging. However, for large driving amplitudes, a substantial fraction of neighbor links do not reform
upon reversal, and the cycle-to-cycle mean square displacements become independent of cycle number: here the motion becomes
diffusive. Our data is consistent with the hypothesis that this strong irreversibility at large cycle amplitudes is connected to the growth of a giant component, which signifies a collective breaking of contacts involving a number of beads that is proportional to the system size. We further showed that the growth of the largest component is consistent with a power law dependence on $f_b$ for the largest reversal amplitude.

We note that our new topological measures of irreversibility --- though less stringent than exact reversal symmetry of the
equations of motion --- correlate well with a strong increase in MSD with cycle amplitude (Fig.~\ref{fig:msdsall}c), and thus are
likely relevant for understanding dynamics of dense granular flows. 

One important future direction of study will be to connect the onset of irreversibility in bead configuration to the emergence of
dynamical heterogeneities.  It is important to note that this work has probed in detail time scales much shorter than those at which dynamical heterogeneities are observed in, for instance,~\cite{PhysRevLett.102.088001}.  As a result, our observations are likely related to the formation of the building blocks of dynamical heterogeneities, whose dynamics may unfold at much larger time scales.

\begin{acknowledgments}
We thank Joost Weijs for his contributions to our particle extraction routine, William Derek Updegraff for his assistance in experimentation, as well as Mark Hererra for his assistance in analysis.  This work was supported by NSF grant DMR0907146.  KR was supported by the Rosztoczy Foundation.
\end{acknowledgments}

\bibliography{granulates2}

\end{document}